\documentstyle[12pt]{article}

\def\MS{{\overline{\hbox{MS}}}}
\def\MSf{{\tiny\overline{\hbox{MS}}}}
\def\pd{\partial}
\def\G{\Gamma}
\def\Ga{\G}

\def\L{\Lambda}
\def\La{\L}
\def\m{\mu}
\def\l{\lambda}
\def\la{\l}

\def\al{\alpha}
\def\th{{\hbox{${1\over3}$}}}
\def\va{\varphi}
\def\h{\hbox{${1\over2}$}}
\def\ha{\h}
\def\pa{\pd}
\def\ar{\rightarrow}
\def\bib{\bibitem}
\def\D{{\cal D}}

\def\intf{\int d^{4}x\,}

\def\hapha{\frac{\hbar}{2(4\pi)^2}}

\def\be{\beta}
\def\ga{\gamma}
\def\de{\delta}

\def\ka{\kappa\,}

\def\beq{\begin{equation}}
\def\eeq{\end{equation}}
\def\bed{\begin{displaymath}}
\def\eed{\end{displaymath}}
\def\beqq{\begin{eqnarray}}
\def\eeqq{\end{eqnarray}}
\def\bedd{\begin{eqnarray*}}
\def\eedd{\end{eqnarray*}}

\advance\hoffset by -.35in
\begin{document}

\begin{flushright}
\bf{DIAS-STP 96-10\\April 1996}
\end{flushright}

\def\sqr#1#2{{\vcenter{\vbox{\hrule height.#2pt
        \hbox{\vrule width.#2pt height#1pt \kern#1pt
           \vrule width.#2pt}
        \hrule height.#2pt}}}}
\def\square{\mathchoice\sqr34\sqr34\sqr{2.1}3\sqr{1.5}3}

\begin{center}\Huge{\bf{A Multi-scale Subtraction Scheme and Partial
Renormalization Group Equations in the $O(N)$-symmetric $\phi^4$-theory}}
\end{center}

\vspace{1cm}

\begin{center} \large{C. Ford \footnote{Present address:
Theor. Phys. Institut, Universit\"at Jena, Fr\"obelsteig 1, D-07743 Jena, 
Germany.  e-mail : Ford@hpfs1.physik.uni-jena.de}
and C. Wiesendanger \footnote{e-mail : wie@stp.dias.ie}\\
School of Theoretical Physics\\
Dublin Institute for Advanced Studies\\
10 Burlington Road\\
Dublin 4, Ireland}
\end{center}

\vspace{1cm}

\begin{abstract}
\small
{To resum large logarithms in multi-scale problems a generalization of
$\MS$ is introduced allowing for as many renormalization scales
as there are generic scales in the problem. In the new \lq\lq minimal
multi-scale subtraction scheme'' standard perturbative boundary conditions
become applicable. However, the multi-loop beta functions depend on the various
renormalization scale ratios and a large logarithms  resummation has to be
performed on them. Using these improved beta functions the \lq\lq partial'' 
renormalization group equations 
corresponding to the renormalization point independence of physical quantities
allows one to resum the logarithms. As an application the leading and
next-to-leading order two-scale analysis of the effective potential in the
$O(N)$-symmetric $\phi^4$-theory is performed. This  calculation
indicates that there is no stable vacuum in the broken phase of the
theory for $1<N\leq4$. }
\end{abstract}

\clearpage

\section{Introduction}

\paragraph{}
The renormalization group (RG) has proved one of the most important tools
in refined perturbative analyses. For it has been recognized for a long time
that ordinary loop-wise perturbation expansions of important physical
quantities
are not only restricted to \lq\lq small'' values of the couplings 
but are often
rendered useless by the occurrence of \lq\lq large'' logarithms.
RG resummation of these logarithms is then crucial to establish a region of
validity for perturbative results.

This is the case in the analysis of vacuum stability (VS) in the Standard Model
(SM), where the loop-expansion of the effective potential (EP) contains
logarithmic terms. Only after RG summation of these logarithms may the
requirement of vacuum stability be turned into bounds on the Higgs
mass \cite{hig1}.
Again, the discussion of Bjorken scaling and its violations in deep inelastic
scattering (DIS) is reliable only after RG summation of the relevant logarithms
yielding in turn high precision tests of QCD and one of the most accurate
determinations of the strong coupling constant \cite{dis1}.

To apply the established RG techniques in both cases it is
essential that in the region of interest (large absolute values of the scalar
field in the discussion of VS, large momentum transfer for
fixed Bjorken variable $x_B$ in DIS) there is only one generic scale
${\cal M}$.
Then, using some mass independent renormalization scheme such as
$\MS$ ${\cal M}$
may be tracked by the corresponding renormalization scale $\mu$, as it occurs
in the combination $\hbar\log({\cal M}/\mu^2)$ only.
Choosing $\mu^2={\cal M}$ removes
the potentially large logarithms from the perturbation series. Hence, at this
scale the perturbative result is trustworthy for \lq\lq small'' values of
the couplings and yields the proper boundary condition for the RG evolution
to finite values of $\hbar\log({\cal M}/\mu^2)$.

However, there may be many generic scales ${\cal M}_i$ in the region of
interest.  For example, in the computation of finite temperature EP
\cite{hight} or in supersymmetric extensions of the SM one encounters
this problem \cite{susy}. But even in the SM there are largely
differing effective scales near the tree-level minimum. Although the usual
VS analyses of the SM were concerned with large absolute values of the scalar
field far away from the tree minimum it is implicitly assumed that the
tree minimum is only slightly shifted by quantum corrections. For
consistency, one should check this assumption; this is a highly non-trivial
multi-scale problem. The breakdown
of the ordinary RG analysis of DIS at small and large $x_B$ is again due to the
growing importance of generic scales other than the large momentum transfer
(for a review see ref. \cite{dis2}). In both instances different potentially
large logarithms
$\hbar\log({\cal M}_i/\mu^2)$ occur in the loop-wise perturbative expansion
which should be resummed in order to get trustworthy results.
But as there is only one renormalization scale one cannot trace the various
${\cal M}_i$ at once and remove all the log's from a loop-wise expansion at
one particular scale. So, although one still has a perfectly good RG equation
there is no longer a proper boundary condition
to RG-evolve from. This problem has been recognized by many authors.

Sticking to the $\MS$ scheme the decoupling theorem \cite{dec} was used
in ref. \cite{bando1} to obtain some regionwise approximation to leading
log's (LL) multi-scale summations. Although this is perfectly reasonable,
one has to employ \lq\lq low-energy'' parameters, and it is not clear  how to
obtain sensible approximations for these low-energy parameters in terms of
the basic parameters of the full theory. Alternatively, one of us \cite{cf2}
argued that one could
still apply the standard $\MS$ RG equation to multi-scale problems
provided \lq\lq improved'' boundary conditions were employed.
Although some improved boundary conditions were suggested in some  
simple cases, no general prescription was given for constructing 
these boundary conditions, and no obvious improved boundary conditions were
apparent for the subleading log's summation. 

Clearly, one must go beyond the usual mass-independent renormalization schemes
if  multi-scale problems are to be seriously tackled.
In the context of the effective potential we are aware of two different
approaches. In ref. \cite{nak} it was argued that one could employ a
mass-dependent scheme in which decoupling of heavy modes is manifest in the
perturbative RG functions.
Alternatively, in ref. \cite{ej} the usual $\MS$ scheme was extended to
include several renormalization scales $\kappa_i$. While this seems to be an
excellent idea, the specific scheme in \cite{ej} has two drawbacks.
Firstly, the number 
of renormalization points does not necessarily match the number
of generic scales in the problem at hand, as there is a RG scale $\ka_i$
associated with each coupling. Secondly, when computing multi-scale RG
functions to $n$ loops one encounters contributions proportional
to $\log^{n-1}(\ka_i/\ka_j)$ (and lower powers). If some of the
$\log(\ka_i/\ka_j)$ are \lq\lq large'' then even the perturbative RG
functions cannot be trusted and used to sum logarithms. A similar approach to
the one of ref. \cite{ej} was outlined in ref. \cite{ni1} though no
detailed perturbative calculations were performed.

In this paper we adopt a more systematic approach. Using the freedom of
\sl finite \rm renormalizations we introduce a new \lq\lq minimal
multi-scale subtraction scheme'' that allows for as many renormalization
scales $\ka_i$ as there are generic scales in the problem. Hence, removing
all large logarithms at scales $\ka_i^2={\cal M}_i$ in the new scheme
standard perturbative boundary conditions become applicable. As in the
approach of ref. \cite{ej}, the multi-loop RG
functions in this scheme \sl inevitably \rm depend on the renormalization scale
ratios $\ka_i/\ka_j$. However, within our minimal multi-scale subtraction
scheme we are able to implement a \sl large logarithms resummation on the RG
functions \rm themselves. Using these improved RG functions the
\lq\lq partial'' RGE's corresponding to the renormalization point independence
of physical quantities allow us then to resum the logarithms for any other
choice of scales. 

Much like the SM, the calculation of the effective potential near the
tree-level minimum of the broken phase ($m^2<0$) in the $O(N)$-symmetric
$\phi^4$-theory is a two-scale problem for $1<N<\infty$. In our opinion,
this is the simplest non-trivial multi-scale problem in four dimensions,
and so we propose to use this model to demonstrate our method. In fact, we are
able to \sl analytically \rm perform leading order (LO) and next-to-leading 
order (NLO) multi-scale computations in the $O(N)$-model. 
Surprisingly, this analysis indicates that the assumption that the tree-level
is not significantly  shifted by quantum corrections is only valid for
$N>4$. For $1<N\leq4$ it appears that there might not even be a stable
vacuum in the broken phase.

The outline 
of the paper is as follows. In section 2 we review the standard $\MS$
RG approach to LL summations in the single-scale cases $N=1$ and $N\ar\infty$.
In section 3 we motivate the idea of two-scale renormalization and introduce
our minimal two-scale subtraction scheme. In section 4 we compute the
leading order two-scale RG functions within our minimal prescription.
We use these LO 
beta functions in section 5 to compute the LO running parameters,
which are then used in section 6 to compute the two-scale RG improved potential
to leading order. In sections 7 and 8 we determine the next-to-leading order
contributions to the RG functions and running parameters. In section 9
we obtain the NLO effective potential. Section 10 is devoted to a discussion
of the special case $N=2$. In appendix A we collect the values of various
constants and in appendix B we discuss some relevant two-loop integrals.

\section{Resumming log's in the effective potential}

\paragraph{}
Let us consider the massive $O(N)$-symmetric field theory with Lagrangian
\beq {\cal L}=\frac{1}{2}\pa_\al\phi\pa^\al\phi
-\frac{\l}{24}\phi^4-\frac{1}{2}m^2\phi^2-\La \eeq
where $\phi$ is an $N$-component scalar field. Note the inclusion of the
cosmological
constant $\La$ \cite{bando2} which will prove essential in the discussion of
the RG later (For a nice discussion of this point in the context of
curved spactime calculations we refer to \cite{buch}).

We are interested here mainly in the effective potential which arises as the
zeroth
order term in a derivative expansion of the effective action $\Ga[\va]$ 
\beq \label{1.2} \Ga[\va]=\intf \left(-V(\va)
+\frac{1}{2}Z(\va)\pa_\al\va\pa^\al\va+{\cal O}(\pa^4)\right). \eeq
As usual $\Ga[\va]$ is the Legendre transform of the Schwinger
functional ${\cal W}[j]$.

A loop-wise perturbation expansion of
$V=\sum\limits_n\frac{\hbar^n}{(4\pi)^{2n}} V^{(\hbox{\tiny{$n$-loop}})}$ 
\cite{cw,ja} 
yields
in the $\MS$-scheme
\beqq \label{1.3} V^{(\hbox{\tiny{tree}})}&=&\frac{\la}{24}\va^4
+\frac{1}{2}m^2\va^2+\La, \nonumber \\
V^{(\hbox{\tiny{$1$-loop}})}&=&\frac{{{\cal M}_1}^2}{4}
\left(\log\frac{{\cal M}_1}{\mu^2}-\frac{3}{2}\right)
+(N-1)\frac{{{\cal M}_2}^2}{4}
\left(\log\frac{{\cal M}_2}{\mu^2}-\frac{3}{2}\right), \eeqq
where
\beq {\cal M}_1=m^2+\h\l\va^2,\quad {\cal M}_2=m^2+\hbox{${1\over6}$}\l\va^2,
\eeq
and $\m$ is the renormalization scale.
The one-loop contribution to the EP thus contains logarithms of
the ratios ${\cal M}_i/\m^2$ to the first power and in general the
$n$-loop contribution will be a polynomial of the $n$th order in these 
logarithms.
(The explicit two-loop result has been obtained in \cite{cf1}.) 
The EP is independent of the renormalization
scale $\mu$ which gives rise to a $\MS$ RG 
equation.

In view of these logarithms the loop-wise expansion may be trusted only
in a region in field- and coupling-space where simultaneously
\beq \frac{\hbar\l}{(4\pi)^2}
\ll 1,\quad \frac{\hbar\l}{(4\pi)^2}\log\frac{{\cal M}_i}{\mu^2}\ll 1, \eeq
conditions which may hardly be fulfilled e.g. around the tree-level minimum
of the potential, where in the broken phase ${\cal M}_2=0$, even with a
judicious choice of
the scale $\mu$. Hence, to obtain a wider range of validity one has to resum
the logarithms in the EP \cite{cw}.

In the two limiting cases $N=1$ and $N\ar\infty$ there is essentially only
one relevant scale involved, ${\cal M}_1$ for $N=1$ and ${\cal M}_2$ for
$N\ar\infty$.
Setting the renormalization scale $\mu$ equal to the relevant scale
removes the potentially large logarithms at this scale and we may trust the
tree-level EP there. To recover the EP at any other
scale we then use the $\MS$ RG
\beq \label{1.6} \D V=0,\quad \D=\mu\frac{\pa}{\pa\mu}+\be_\la\frac{\pa}
{\pa\la}
+\be_{m^2}\frac{\pa}{\pa m^2}+\be_\La\frac{\pa}{\pa\La}
-\be_\va\va\frac{\pa}{\pa\va}. \eeq
We next expand the 
RG functions in powers of $\hbar$. As the expansion coefficient
$Z(\va)$ in (\ref{1.2}) does not contain logarithms at the one-loop level no
anomalous field-dimension arises and it is an easy task to read off the other
one-loop coefficients from the result (\ref{1.3}). For $N=1$ we have 
at \sl one loop \rm
\beq \label{1.7} _1\be_\la=\frac{3\hbar\la^2}{(4\pi)^2},
\quad _1\be_{m^2}=\frac{\hbar\l m^2}{(4\pi)^2},\quad 
{}_1\be_\La=\frac{\hbar m^4}{2(4\pi)^2} ,\quad _1\be_\va=0, \eeq
whereas for $N\ar\infty$ we find
\beq \label{1.8} _2\be_\la=\frac{\hbar N\l^2}{3(4\pi)^2}
,\quad _2\be_{m^2}=\frac{\hbar N\l m^2}{3(4\pi)^2},\quad 
_2\be_\La=\frac{\hbar Nm^4}{2(4\pi)^2}
,\quad _2\be_\va=0 \eeq
which are exact in this limit.

With the use of the RG functions we next recover the running couplings. Setting
$s=\frac{\hbar}{(4\pi)^2}\log (\m(s)/\m)$, where $\mu$ is the reference scale,
 we
have for $N=1$
\beqq & &\la(s)=\la(1-3\la s)^{-1},\quad m^2(s)
=m^2(1-3\la s)^{-1/3} \nonumber \\
& &\La(s)=\la-\frac{m^4}{2\la}\left[(1-3\la 
s)^{1/3}-1\right] \eeqq
and for $N\ar\infty$
\beqq & &\la(s)=\la(1-\th N\la s)^{-1}
,\quad m^2(s)=m^2(1-\th N\la s)^{-1} \nonumber \\
& &\La(s)=\La+\frac{3m^4}{2\la}
\left[(1-\th N\la s)^{-1}-1\right]. \eeqq

Imposing the tree-level boundary condition the LL approximation to the
 effective
potential at an arbitrary scale $\mu$ becomes
\beq \label{1.11} V_i^{(0)}
\bigl(\la,m^2,\va,\La;\mu\bigr)=\frac{\la(s_i)}{24}\va^4
+\frac{1}{2}m^2(s_i)\va^2+\La(s_i) \eeq
where
\beq s_1=\hapha\log\frac{{\cal M}_1}{\mu^2},
\quad s_2=\hapha\log\frac{{\cal M}_2}{\mu^2}. \eeq
Higher orders may again be systematically
resummed giving rise to the NLL, NNLL, ...
approximations to the effective potential \cite{kast}.

As the usual RG may cope with one scale only this approach does not allow  a
systematic resummation in the generic case as we have to deal with two relevant
scales, at least near 
the tree-level minimum in the broken phase. Therefore, we have
to generalize the usual 
RG approach allowing for as many renormalization scales as
there are relevant scales 
in the theory, the task we turn to in the next section.

\section{Two-scale renormalization}

\paragraph{}
In the previous 
section we were able to use the renormalization scale $\mu$ arising
in $\MS$ to track one relevant scale and to resum the corresponding
logarithms with the $\MS$ RG. This was sufficient to obtain a trustworthy
approximation to the 
EP for $N=1$ and $N\ar\infty$. To deal with the general case
we shall introduce
a new set of parameters depending on \sl two \rm renormalization scales
$\ka_1,\ka_2$
which allow us to track the two generic scales ${\cal M}_i$. That is, we
consider a \sl finite \rm transformation
\beqq \label{2.13} \la_\MSf&=&F_\la(\la;\ka_1,\ka_2,\mu) \nonumber \\
m^2_\MSf&=&m^2 F_{m^2}(\la;\ka_1,\ka_2,\mu) \nonumber \\
\La_\MSf&=&\La+m^4 F_\La(\la;\ka_1,\ka_2,\mu) \nonumber \\
\va_\MSf&=&\va F_\va(\la;\ka_1,\ka_2,\mu). \eeqq
Here, the $\MS$ parameters 
$\la_\MSf,m^2_\MSf,\La_\MSf,\va_\MSf$ at scale $\mu$ 
may be regarded as \lq\lq bare'' ones as opposed
to the new \lq\lq renormalized'' two-scale subtraction scheme parameters
$\la,m^2,\La,\va$.

Our goal is to construct a transformation (\ref{2.13}) with the 
following properties:

i) The effective action $\Ga$, when expressed in terms of the new parameters,
should be independent of the $\MS$ scale $\mu$.

ii) When $\ka_1=\ka_2$ the minimal two-scale subtraction scheme should
coincide with $\MS$ at that scale.

iii) When $N=1$ or $N\ar\infty$ one scale should drop and
the  two-scale  scheme should coincide with $\MS$
at the remaining scale.

iv) When $\ka_i^2={\cal M}_i$ the standard loop-expansion should
render a reliable approximation to the
full EP insofar as ${\hbar\over {(4\pi)^2}}\la(\ka_1,\ka_2)$ is \lq\lq small''.

In order to find 
a suitable transformations (\ref{2.13}) with the desired properties
we first study the 
associated RG's and RG functions. Having obtained a trusworthy
set of RG functions we turn them into running couplings and an improved
effective potential.

Our starting point is
\beq \Ga_\MSf\bigl[\la_\MSf,m^2_\MSf,\La_\MSf,\va_\MSf;\mu\bigr]
=\Ga[\la,m^2,\La,\va;\ka_1,\ka_2] \eeq
from which we derive the 
two RGE's corresponding to variations of scales $\ka_i$,
where the other scale 
$\ka_j$ and the $\MS$ parameters are held fixed, in much the
same way as the $\MS$ RG is usually derived. Specializing to the effective
potential we obtain
\beq \label{2.15}
\D_i V=0,\quad \D_i=\ka_i\frac{\pa}{\pa\ka_i}+{_i\be}_\la\frac{\pa}{\pa\la}
+{_i\be}_{m^2}\frac{\pa}{\pa m^2}+{_i\be}_\La\frac{\pa}{\pa\La}
-{_i\be}_\va\va\frac{\pa}{\pa\va}. \eeq
The two sets of RG functions are defined as usual
\beq \label{2.16} {_i\be}_\la=\ka_i\frac{d\la}{d\ka_i},\quad
{_i\be}_{m^2}=\ka_i\frac{d m^2}{d\ka_i},\quad
{_i\be}_\La=\ka_i\frac{d\La}{d\ka_i},\quad
{_i\be}_\va\va=-\ka_i\frac{d\va}{d\ka_i} \eeq
for $i=1,2$. In general they may be functions not only of $\la,m^2$ as are the
$\MS$ RG functions but also of $\ka_2/\ka_1$.

Note that property ii) requires that the sum of the two-scale RG functions at
$\ka_1=\ka_2$ coincides with the $\MS$ RG function at that scale
\beq _1\be_{.}(\ka_1=\ka_2)+_2\be_{.}(\ka_1=\ka_2)=\be_{.\, ,\MSf} \eeq
and  property iii) fixes the two sets of RG functions in the
single-scale limits. For $N=1$ there are no Goldstone bosons. Hence, we
have to choose the usual $N=1$ $\MS$ RG functions as the first set
of RG functions, given to ${\cal O}(\hbar)$ by eqn. (\ref{1.7}), and to
disregard the second set of RG functions so that $\D_2=\ka_2\pa/\pa\ka_2$.
For $N\ar\infty$ there are no Higgs contributions. Accordingly, in this
limit we have to disregard the first set of RG functions,
so that $\D_1=\ka_1\pa/\pa\ka_1$,
and to choose the second set as the large $N$ $\MS$ RG functions, given by
eqn. (\ref{1.8}).

Let us come back to the general case. As we want to vary $\ka_1$ and $\ka_2$
independently we must respect the integrability conditions
\beq \label{2.18} [\ka_1d/d\ka_1,\ka_2d/d\ka_2]=[\D_1,\D_2]=0. \eeq
An essential feature of a mass-independent renormalization scheme such as
$\MS$ is that the beta functions do not depend on the renormalization scale
$\mu$. Unfortunately we cannot generalize this to the multi-scale case
and demand that the two sets of beta functions be independent of $\ka_2/\ka_1$.
The point is that the independence of the RG functions from the
scales $\ka_i$, ie. $[\ka_i\pd/\pd\ka_i,\D_j]=0$, is incompatible with the 
integrability condition eqn. (\ref{2.18}). However, it is
still possible to arrange for one of the two sets of RG functions, or in slight
generalization for a linear combination of the two sets, to be
$\ka_i$-independent. Hence, we assume that
\beqq & &{\tilde\be}_\la={_1\be}_\la p_1+{_2\be}_\la p_2
,\quad {\tilde\be}_{m^2}={_1\be}_{m^2} p_1+{_2\be}_{m^2} p_2 \nonumber \\
& &{\tilde\be}_\La={_1\be}_\La p_1+{_2\be}_\La p_2
,\quad {\tilde\be}_\va={_1\be}_\va p_1+{_2\be}_\va p_2 \eeqq
depend only on $\la,m^2$ unlike the RG functions ${_i\be}_{.}$ in (\ref{2.16}).
Accordingly,
their values in a perturbative expansion may be trusted for small $\la$
whatever the value of $\ka_2/\ka_1$. $p_j$ are real numbers subject
to $p_1+p_2=1$. The corresponding RG operator
\beqq {\tilde\D}&=&p_1\D_1+p_2\D_2 \nonumber \\
&=&p_1\ka_1\frac{\pa}{\pa\ka_1}+p_2\ka_2\frac{\pa}{\pa\ka_2}+
{\tilde\be}_\la\frac{\pa}{\pa\la}
+{\tilde\be}_{m^2}\frac{\pa}{\pa m^2}+{\tilde\be}_\La\frac{\pa}{\pa\La}
-{\tilde\be}_\va\va\frac{\pa}{\pa\va} \eeqq
commutes with $\ka_i{\pa}/{\pa\ka_i}$. To recover the 
${\ka_2}/{\ka_1}$-dependence
of $\D_i$ we use that eqn. (\ref{2.18}) implies
\beq \label{2.21} [{\tilde\D},\D_i]=0, \eeq
yielding RG-type equations for the sought-after ${_i\be}_{.}$. 
We remark that the final \lq\lq improved'' potential will have a strong
dependence on the $p_j$-parameters. Each $p_j$-choice corresponds to a 
different transformation in eqn. (\ref{2.13}) which satisfies
conditions i), ii) and iii). Accordingly, we should decide for which
values of $p_j$ the transformation (\ref{2.13}) \lq\lq best'' meets
condition iv). In section 6 we will argue that the appropriate 
choice is $p_1=1$ and $p_2=0$. That is, the first set of beta
functions, which track the Higgs scale, are independent of $\ka_2/\ka_1$.

\section{LO RG functions}

\paragraph{}
To determine the ${_i\be}_{.}$ we make a perturbative ansatz
\beq \label{4.1}{_i\be}_{.}(\la,m^2;t)=\sum_{a=0}^\infty {\hbar^{a+1}\over
{(4\pi)^{2a+2}}}\,{_i\be}^{(a)}_{.}(\la,m^2;t)
,\quad t={\hbar\la\over{(4\pi)^2}} \log\frac{\ka_2}{\ka_1}. \eeq
Note that this is \sl not \rm simply a loop-expansion, since 
although we expand in $\hbar$ we retain all orders in $t$. Rather,
we should view eqn. (\ref{4.1}) as a LL, NLL, ... expansion of the
two-scale RG functions.
Hence, we assume the full $\ka_i$-dependence of $_i\be_{.}$ to enter 
via $t$. This immediately allows us to rewrite
\beq p_1\ka_1\frac{\pa}{\pa\ka_1}+p_2\ka_2\frac{\pa}{\pa\ka_2}=
{\hbar\over{(4\pi)^2}}\la (p_2-p_1)\frac{\pa}{\pa t}
\equiv{\hbar\over{(4\pi)^2}}D. \eeq
The corresponding perturbative decomposition of the RG operators becomes
\beqq & &\D_i=\sum_{a=0}^\infty {\hbar^{a+1}\over
{(4\pi)^{2a+2}}}\,\D_i^{(a)} \nonumber \\
& &\D_i^{(a)}={(4\pi)^2\over{\hbar}}
\de^{a0}\ka_i\frac{\pa}{\pa\ka_i}+{_i\be}^{(a)}_\la\frac{\pa}{\pa\la}
+{_i\be}^{(a)}_{m^2}\frac{\pa}{\pa m^2}+{_i\be}^{(a)}_\La\frac{\pa}{\pa\La}
-{_i\be}_\va^{(a)}\va\frac{\pa}{\pa\va} \eeqq
with analogous expressions for ${\tilde\D},{\tilde\D}^{(a)}$. To determine
the respective RG-like equation for a given order ${_i\be}^{(a)}_{.}$ we need
\beqq \label{2.25} [{\tilde\D}^{(a)},\D_i^{(b)}]
&=&\left(\de^{a0}D\,{_i\be}^{(b)}_\la
+{\tilde\be}^{(a)}_\la\frac{\pa}{\pa\la} {_i\be^{(b)}_\la}
-{_i\be}^{(b)}_\la\frac{\pa}{\pa\la}{\tilde\be}^{(a)}_\la
\right)\frac{\pa}{\pa\la} \nonumber \\
&+&\left(\de^{a0}D\,{_i\be}^{(b)}_{m^2}
+{\tilde\be}^{(a)}_\la\frac{\pa}{\pa\la} {_i\be^{(b)}_{m^2}}
-{_i\be}^{(b)}_\la\frac{\pa}{\pa\la}{\tilde\be}^{(a)}_{m^2}
\right)\frac{\pa}{\pa{m^2}} \nonumber \\
&+&\left(\de^{a0}D\,{_i\be}^{(b)}_\La
+{\tilde\be}^{(a)}_\la\frac{\pa}{\pa\la} {_i\be^{(b)}_\La}
-{_i\be}^{(b)}_\la\frac{\pa}{\pa\la}{\tilde\be}^{(a)}_\La\right. \nonumber \\
& &\left.\,\,\,+{\tilde\be}^{(a)}_{m^2}\frac{\pa}{\pa m^2} {_i\be^{(b)}_\La}
-{_i\be}^{(b)}_{m^2}\frac{\pa}{\pa m^2}{\tilde\be}^{(a)}_\La
\right)\frac{\pa}{\pa\La} \nonumber \\
&-&\left(\de^{a0}D\,{_i\be}_\va^{(b)}
+{\tilde\be}^{(a)}_\la\frac{\pa}{\pa\la} {_i\be_\va^{(b)}}
-{_i\be}^{(b)}_\la\frac{\pa}{\pa\la}{\tilde\be}_\va^{(a)}
\right)\va\frac{\pa}{\pa\va}. \eeqq
Here, we used the form of eqn. (\ref{2.13}) implying, in generalization of the
single-scale case, that ${_i\be}_\la,{_i\be}_\va$ do not depend on
$m^2,\La$ and ${_i\be}_{m^2},{_i\be}_\La$ not on $\La$.
We can write
\beqq \label{2.26} & &{_i\be}_\la^{(a)}=\la^{a+2}\al_i^{(a)}(t)
,\quad {_i\be}_{m^2}^{(a)}=m^2\la^{a+1}\be_i^{(a)}(t) \nonumber \\
& &{_i\be}_\La^{(a)}=m^4\la^{a}\ga_i^{(a)}(t)
,\quad {_i\be}_\va^{(a)}=\la^{a+1}\de_i^{(a)}(t) \eeqq
with analogous but $t$-independent expressions for the
$\tilde\be^{(a)}_.$.

At LO we have $a=b=0$ and eqn. (\ref{2.21}) reduces to
\beq [{\tilde\D}^{(0)},\D_i^{(0)}]=0. \eeq
The corresponding equations for the various ${_i\be}^{(0)}_{.}$ may be read
off from eqn.
(\ref{2.25}). We now solve them in turn.

${_i\be}^{(0)}_\la$ is determined by
\beq D\,{_i\be}^{(0)}_\la
+{\tilde\be}^{(0)}_\la\frac{\pa}{\pa\la} {_i\be^{(0)}_\la}
-{_i\be}^{(0)}_\la\frac{\pa}{\pa\la}{\tilde\be}^{(0)}_\la=0. \eeq
Inserting the further decomposition (\ref{2.26}) and taking into account
the $\la$-dependence of $t$ this equation reduces to
\beq \left(p_2-p_1+{\tilde\al}^{(0)}t\right)\frac{\pa}{\pa t}\al_i^{(0)}=0. \eeq
Hence, $\al_i^{(0)}$ is independent of $t$
\beqq \al_i^{(0)}(t)&=&a_{1i}^{(0)}=\al_i^{(0)}(0), \nonumber \\
{_i\be}_\la^{(0)}&=&\la^{2}\al_i^{(0)}. \eeqq

The equation for ${_i\be}^{(0)}_{m^2}$ is
\beq D\,{_i\be}^{(0)}_{m^2}
+{\tilde\be}^{(0)}_\la\frac{\pa}{\pa\la} {_i\be^{(0)}_{m^2}}
-{_i\be}^{(0)}_\la\frac{\pa}{\pa\la}{\tilde\be}^{(0)}_{m^2}=0 \eeq
and reduces to 
\beq \left(p_2-p_1+{\tilde\al}^{(0)}t\right)\frac{\pa}{\pa t}\be_i^{(0)}
+{\tilde\al}^{(0)}\be_i^{(0)}=\be_i^{(0)}{\tilde\al}^{(0)}. \eeq
Its solution is best expressed in terms of the function $f$
\beq \label{4.12} f(t)\equiv\frac{p_2-p_1+{\tilde\al}^{(0)}t}{p_2-p_1} \eeq
and reads
\beqq \label{3.34}
& &\be_i^{(0)}(t)=b_{1i}^{(0)}+b_{2i}^{(0)} f^{-1}(t), \nonumber \\
& &{_i\be}_{m^2}^{(0)}=m^2\la\be_i^{(0)}, \eeqq
where
\beqq 
& &b_{1i}^{(0)}={\tilde B}^{(0)}a_{1i}^{(0)} \quad\mbox{and}\quad 
{\tilde B}^{(0)}=\frac{{\tilde\be}^{(0)}}{{\tilde\al}^{(0)}}, \nonumber \\
& &b_{2i}^{(0)}=
\frac{1}{{\tilde\al}^{(0)}}\left({\tilde\al}^{(0)}\be_i^{(0)}(0)-
\al_i^{(0)}(0){\tilde\be}^{(0)}\right). \eeqq

The determination of ${_i\be}^{(0)}_\La$ is a bit more involved
\beq D\,{_i\be}^{(0)}_\La
+{\tilde\be}^{(0)}_\la\frac{\pa}{\pa\la} {_i\be^{(0)}_\La}
-{_i\be}^{(0)}_\la\frac{\pa}{\pa\la}{\tilde\be}^{(0)}_\La 
+{\tilde\be}^{(0)}_{m^2}\frac{\pa}{\pa m^2} {_i\be^{(0)}_\La}
-{_i\be}^{(0)}_{m^2}\frac{\pa}{\pa m^2}{\tilde\be}^{(0)}_\La=0. \eeq
The corresponding reduced ODE then reads
\beq \left(p_2-p_1+{\tilde\al}^{(0)}t\right)\frac{\pa}{\pa t}\ga_i^{(0)}
+2{\tilde\be}^{(0)}\ga_i^{(0)}=2\be_i^{(0)}{\tilde\ga}^{(0)} \eeq
and is solved by
\beqq \label{3.37}
\ga_i^{(0)}(t)&=&c_{1i}^{(0)}+c_{2i}^{(0)} f^{-1}(t)
+c_{3i}^{(0)}f^{-2{\tilde B}^{(0)}}(t), \nonumber \\
{_i\be}_\La^{(0)}&=&m^4\ga_i^{(0)}, \eeqq
where
\beqq
& &c_{1i}^{(0)}={\tilde C}^{(0)}a_{1i}^{(0)} \quad\mbox{and}\quad 
{\tilde C}^{(0)}=\frac{{\tilde\ga}^{(0)}}{{\tilde\al}^{(0)}}, \quad
c_{2i}^{(0)}=\frac{2{\tilde C}^{(0)}}{2{\tilde B}^{(0)}-1}
b_{2i}^{(0)}, \nonumber \\
& &c_{3i}^{(0)}=\frac{1}{{\tilde\al}^{(0)}}
\left({\tilde\al}^{(0)}\ga_i^{(0)}(0)-
\al_i^{(0)}(0){\tilde\ga}^{(0)}\right)-c_{2i}^{(0)}. \eeqq

As for ${_i\be}^{(0)}_\va$ the trivial boundary condition (see below) implies
\beq {_i\be}^{(0)}_\va=0. \eeq

In this section we have computed the two-scale LO RG functions for 
the $O(N)$-model. The results depend on $p_j$ as well as the 
boundary conditions $\al_i^{(0)}(0)$, $\be_i^{(0)}(0)$,
$\ga_i^{(0)}(0)$, 
$\delta_i^{(0)}(0)$ which determine the RG functions at $t=0$
(ie. $\ka_1=\ka_2$). In fact, at LO the boundary conditions are
\sl uniquely \rm determined by the single-scale limit conditions
following from  requirements ii) and iii) in section 3
\bed \al_1^{(0)}(0)=3,\quad\!\!\!
\be_1^{(0)}(0)=1, \quad\!\!\! 
\ga_1^{(0)}(0)=\h,\quad\!\!\!
\delta_1^{(0)}(0)=0, \eed
\beq \al_2^{(0)}(0)=\th({N-1}),\!\!\!\quad
\be_2^{(0)}(0)=\th(N-1),\!\!\!\quad
\ga_2^{(0)}(0)=\ha(N-1),\!\!\!\quad
\delta_2^{(0)}(0)=0.\eeq

The LO RG functions 
for $\l$ and $\va$ are independent of $p_j$, and are given by
(some relevant constants are given in appendix A)
\beq
{}_1\be_\la^{(0)}=3\la^2,\quad
{}_2\be_\la^{(0)}=\th(N-1)\la^2,\quad
{}_1\be_\va^{(0)}={}_2\be_\va^{(0)}=0. \eeq
However, the LO RG functions for $m^2$ and $\La$ still have a marked dependence
on $p_j$. As mentioned in the previous section, we are eventually going
to adopt the choice $p_1=1$, $p_2=0$. For this choice eqns. (\ref{3.34}) and
(\ref{3.37}) reduce to
\beq
{}_1\be_{m^2}^{(0)}=m^2\la,\quad\!\!\!
{}_2\be_{m^2}^{(0)}=(N-1)\left[
\hbox{${1\over9}$}+\hbox{${2\over9}$}(1-3t)^{-1}\right]m^2\la \eeq
and
\beq
{}_1\be_\La^{(0)}=\ha m^4,\quad\!\!\!
{}_2\be_\La^{(0)}=(N-1)\left[
\hbox{${1\over{18}}$}-\hbox{${2\over9}$}(1-3t)^{-1}+
\hbox{${2\over3}$}(1-3t)^{-\frac{2}{3}}\right]m^4, \eeq
respectively.
It is clear that \sl the beta functions possess Landau poles \rm at $3t=1$.
Thus, these beta functions are only trustworthy for $1\gg 3t$. 
Returning to the general $p_j$-case, the beta functions have a
Landau pole at $p_1-p_2=\tilde\alpha^{(0)}t$. To avoid this pole we require
$p_1-p_2\gg\tilde\alpha^{(0)}t$ for $p_1>p_2$ and
$p_1-p_2\ll\tilde\alpha^{(0)}t$ for $p_1<p_2$.
The case $p_1=p_2=\ha$ appears to be pathological.

\section{LO running two-scale parameters}

\paragraph{}
The running parameters in the minimal two-scale subtraction scheme
are functions of the variables
\beq \label{5.1} s_i={\hbar\over{(4\pi)^2}}\log\frac{\ka_i(s_i)}{\ka_i},\quad
t={\hbar\la\over{(4\pi)^2}}\log\frac{\ka_2}{\ka_1}, \eeq
where $\ka_i$ are the reference scales. Note that $t(s_i)$ as given in
eqn. (\ref{4.1}) is in fact $s_i$-dependent,
$t(s_i)={\hbar\la(s_i)\over{(4\pi)^2}}\log\frac{\ka_2(s_2)}{\ka_1(s_1)}$.
The running coupling may be expanded in a series in $\hbar$
\beq  \label{5.2} \la(s_i,t)
=\sum_{a=0}^\infty {\hbar^a\over{(4\pi)^{2a}}}\,\la^{(a)}(s_i,t) \eeq
with analogous expansions for $m^2(s_i,t),\La(s_i,t),\va(s_i,t)$.
We now insert these expansions into eqn. (\ref{2.16}) and solve for the
LO parameters.

The equation for the leading order running two-scale coupling is
\beq \frac{d\la^{(0)}}{d s_i}={\la^{(0)}}^2\al_i^{(0)}. \eeq
As $\al_i^{(0)}$ is constant it is easily integrated
\beq \label{4.42} \la^{(0)}(s_i)=\la\left(1-\la(\al_1^{(0)}s_1+
\al_2^{(0)}s_2)\right)^{-1}, \eeq
where the boundary condition is $\la(s_i=0)=\la$.

Turning to the running mass we have to solve
\beq \label{4.44} \frac{d {m^2}^{(0)}}{d s_i}
={m^2}^{(0)}\la^{(0)}\be_i^{(0)}. \eeq
$\be_i^{(0)}$ is given in eqn. (\ref{3.34}) in terms of the function $f(t)$.
As to leading order 
\beq \label{5.6} t(s_i)=\la^{(0)}(s_i)\left(s_2-s_1+\frac{t}{\la}\right) \eeq
the $s_i$-dependence of the r.h.s. of eqn. (\ref{4.44}) is quite involved.
Its integration yields
\beq \label{4.46} {m^2}^{(0)}(s_i)=m^2
\left(\frac{\la^{(0)}(s_i)}{\la}\right)^{B^{(0)}}
\left(\frac{f^{(0)}(s_i)}{f}\right)^{{\tilde B}^{(0)}-B^{(0)}}, \eeq
with
$B^{(0)}=\frac{\be_1^{(0)}+\be_2^{(0)}}{\al_1^{(0)}+\al_2^{(0)}}$, 
and with the boundary condition $m^2(s_i=0)=m^2$.
Here, $f^{(0)}(s_i)$ is the function obtained by inserting eqn. (\ref{5.6})
into eqn. (\ref{4.12}) defining $f(t)$
\beq f^{(0)}(s_i)
=\frac{\la^{(0)}(s_i)}{\la}
\left(1+\frac{(\al_1^{(0)}+\al_2^{(0)})\la(p_1 s_2-p_2 s_1)
+{\tilde\al}^{(0)}t}{p_2-p_1}\right) \eeq
and $f=f^{(0)}(s_i=0)$. Note that if the two scales coincide we have
$t=0$ and $f^{(0)}(s_1=s_2)=f=1$. Requirement ii) provides us then with
a \sl strong check \rm on the correct algebra for the running 
LO and NLO parameters.

We finally determine the running cosmological constant from
\beq \frac{d \La^{(0)}}{d s_i}
=({m^2}^{(0)})^2\ga_i^{(0)}. \eeq
With the use of the results (\ref{4.46}) for ${m^2}^{(0)}$ and (\ref{3.37})
for $\ga_i^{(0)}$ we obtain
\beqq \label{4.49}
\La^{(0)}(s_i)&=&\La+L_1^{(0)}
\left[\frac{({m^2}^{(0)}(s_i))^2}{\la^{(0)}(s_i)}-\frac{m^4}{\la}\right]
\nonumber \\
&+&L_2^{(0)}\left[\frac{({m^2}^{(0)}(s_i))^2}{\la^{(0)}(s_i)}
f^{(0)}(s_i)^{1-2{\tilde B}^{(0)}}-
\frac{m^4}{\la}f^{1-2{\tilde B}^{(0)}}\right], 
\eeqq
where
\beq
L_1^{(0)}=\frac{{\tilde C}^{(0)}}{2{\tilde B}^{(0)}-1}, \quad
L_2^{(0)}=\frac{C^{(0)}}{2B^{(0)}-1}-L_1^{(0)}, \eeq
with $C^{(0)}=\frac{\ga_1^{(0)}+\ga_2^{(0)}}{\al_1^{(0)}+\al_2^{(0)}}$,
and $\La(s_i=0)=\La$.

To LO there is no anomalous field dimension, and so the field parameter
$\va$ does not run.

The LO running coupling $\la^{(0)}(s_i)$ has 
a Landau pole at $\la(\al_1^{(0)}s_1+\al_2^{(0)}s_2)=1$ and 
clearly our approximation will break down before this pole is 
reached. If we let one of the $s_i\ar-\infty$ (ie. the far 
IR region) while holding the other fixed the coupling will 
tend to zero as $\la^{(0)}(s_i\ar-\infty)\propto (-s_i)^{-1}$.
Also note that the LO running coupling is independent
of $p_j$ which parameterize the class of finite renormalizations 
under investigation.  

The behaviour of the running mass and cosmological constant is more 
complicated. Consider the combination
\beq \left(\frac{f^{(0)}(s_i)}{f}\right)
\left(\frac{\la^{(0)}(s_i)}{\la}\right)^{-1}
=1+\frac{(\al_1^{(0)}+\al_2^{(0)})\la(p_1 s_2-p_2 s_1)}
{p_2-p_1+{\tilde\al}^{(0)}t}. \eeq
In the limit investigated $\frac{f^{(0)}(s_i)}{f}$ is not generally
positive unless $p_1=0$ or $p_1=1$. Of course, we thereby assume that
$t$ is chosen such as to avoid the beta function poles in which case
$p_2-p_1+{\tilde\al}^{(0)}t$ has the same sign as $p_2-p_1$.
This is disturbing because in eqns. (\ref{4.46}) and (\ref{4.49})
we are required to take 
non-integer powers of this quantity. Thus, unless $p_1=0$ or $p_1=1$
we are faced with the disquieting possibility of \sl complex \rm
running $m^2$ and $\L$ in a region where the running coupling is 
very small. Fortunately, we will see in the next section that a comparison of 
our $p_j$-dependent improved potential with standard two-loop and 
next-to-large $N$ calculations indicates that $p_1=1$ is the
\lq\lq natural'' choice.

\section{LO RG improved potential}

\paragraph{}
It is now an easy task to turn the results
for the running two-scale parameters into a RG improved effective
potential. Eqn. (\ref{2.15}) yields the identity
\beq \label{6.1}
V(\la,m^2,\va,\La;\ka_1,\ka_2)=
V\bigl(\l(s_i),m^2(s_i),\va(s_i),\Lambda(s_i);\ka_1(s_1),\ka_2(s_2)),
\eeq
with $\ka_i(s_i)$ defined in (\ref{5.1}).
Next, we assume the validity of condition iv) in section 3. Hence, if
\beq \label{6.2} \ka_i(s_i)^2={\cal M}_i(s_j)
\equiv m^2(s_j)+k_i\,\la(s_j)\va^2(s_j), \quad\quad
k_1={1\over 2},\;k_2={1\over 6} \eeq
the loop-expansion of the EP should render
a reliable approximation to the RHS of eqn. (\ref{6.1}).

To proceed we have to determine the values of $s_i$ fulfilling (\ref{6.2}).
Insertion of the $\ka_i(s_i)^2$ from (\ref{6.2}) into (\ref{5.1}) yields a quite
implicit set of equations
\beq \label{6.3} s_i={\hbar\over{2(4\pi)^2}}
\log\frac{{\cal M}_i(s_j)}{\ka_i^2}. \eeq
Since we are meant to be summing consistently all logarithms we have to
solve (\ref{6.3}) iteratively
\beq \label{6.4} s_i=\sum_{a=0}^\infty {\hbar^{a}\over{(4\pi)^{2a}}}
\,s_i^{(a)}(\la,...;s_i^{(0)}) \eeq
in terms of the LO log's
\beq s_i^{(0)}={\hbar\over{2(4\pi)^2}}\log\frac{{\cal M}_i}{\ka_i^2},
\quad\mbox{where}\quad {\cal M}_i={\cal M}_i(s_j=0). \eeq
This yields contributions to the $s_i^{(a)}$ from both the $s_i$-dependence
of the running two-scale parameters and from their own $\hbar$
expansion (\ref{5.2}). For later use we also give the NLO term of the result
\beqq s_i^{(1)}(\la,...;s_i^{(0)})&=&
\ha\log\frac{{\cal M}^{(0)}_i(s_i^{(0)})}{{\cal M}_i},
\quad\mbox{where}\quad \nonumber \\
{\cal M}^{(0)}_i(s_j)&=&{m^2}^{(0)}(s_j)+k_i \la^{(0)}(s_j)\va^2. \eeqq

To obtain the corresponding series expansion for the RG improved effective
potential
\beq \label{6.7} V(\la,...;\ka_i)
=\sum_{a=0}^\infty {\hbar^{a}\over{(4\pi)^{2a}}}\,V^{(a)}(\la,...;\ka_i) \eeq
we approximate the RHS of eqn. (\ref{6.1}) with those terms in the minimal
two-scale subtraction scheme result for the EP surviving when
$\ka_i(s_i)^2={\cal M}_i(s_j)$. 
To ${\cal O}(\hbar)$ they are explicitly given by
\beqq \label{6.8} V\bigl(\l(s_i),...;{\cal M}_i(s_j))
&=&{\l(s_i)\over{24}}
\va(s_i)^4+{1\over2}{m^2}(s_i)\va(s_i)^2+\L(s_i) \nonumber \\
&-&{3\hbar\over{2(4\pi)^2}}\left(\frac{{{\cal M}_1(s_i)}^2}{4}
+(N-1)\frac{{{\cal M}_2(s_i)}^2}{4}\right). \eeqq
We finally insert the running two-scale parameters from (\ref{5.2})
into the RHS of (\ref{6.8}) with their arguments $s_i$ coming from (\ref{6.4}).
Accordingly, an expansion in powers of $\hbar$ yields contributions to the
$V^{(a)}$ from both the $s_i$-dependence of the running two-scale parameters
and from their own $\hbar$-expansion. Keeping only leading order terms we
obtain the LO two-scale RG improved effective potential in the minimal
two-scale subtraction scheme
\beq \label{6.9} V^{(0)}(\la,...;\ka_i)={\l^{(0)}(s_i^{(0)})\over{24}}\va^4
+{1\over2}{m^2}^{(0)}(s_i^{(0)})\va^2+\L^{(0)}(s_i^{(0)}). \eeq

Let us next examine its properties. In the single-scale limits $N=1$
and $N\rightarrow\infty$ eqn. (\ref{6.9}) reduces to eqn. (\ref{1.11})
for $i=1$ and $i=2$, respectively. In the general case $1<N<\infty$
the $m^2\va^2$- and $\Lambda$-terms in eqn. (\ref{6.9}) depend on $p_j$
which parameterizes the class of finite renormalizations under consideration.
Comparison with two-loop and next-to-large $N$  results will provide
us now with a natural value for them.  

We have used a \sl two-scale \rm RG to track the two scales ${\cal M}_1$
and ${\cal M}_2$. Once the two log's have been summed up we can set
$\ka_1=\ka_2=\mu$, ie. we may write our improved potential in standard
$\MS$ parameters. In this way we can compare the improved potential (\ref{6.9})
with standard perturbation theory. When now inserting the various constants and
expanding eqn. (\ref{6.9}) in $s_i^{(0)}$ up to second order
\beqq \label{6.11} V^{(0)}(\la,m^2,\va,\La;\mu)&=&
{\la\over{24}}\va^4\Biggl[
1+3\la s_1^{(0)}+\frac{N-1}{3}\la s_2^{(0)}+9\la^2 {s_1^{(0)}}^2 \nonumber \\
&&+2(N-1)\la^2 s_1^{(0)}s_2^{(0)}
+\frac{(N-1)^2}{9}\la^2 {s_2^{(0)}}^2\Biggr] \nonumber \\
&+&\ha m^2\va^2\Biggl[
1+\la s_1^{(0)}+\frac{N-1}{3}\la s_2^{(0)} \nonumber \\
&&+\left(2-\frac{(N-1)p_2}{3(p_2-p_1)}\right)\la^2 {s_1^{(0)}}^2
+\frac{N-1}{3}\left(2+\frac{2p_2}{p_2-p_1}\right)
\la^2 s_1^{(0)}s_2^{(0)} \nonumber \\
&&+\frac{N-1}{9}
\left(N+2-\frac{3p_2}{p_2-p_1}\right)\la^2 {s_2^{(0)}}^2\Biggr] \nonumber \\
&+&{m^4\over{\la}}\Biggl[
\ha\la s_1^{(0)}+\frac{N-1}{2}\la s_2^{(0)}
+\left(\ha-\frac{(N-1)p_2}{3(p_2-p_1)}\right)\la^2 {s_1^{(0)}}^2 \nonumber \\
&&+\frac{N-1}{3}
\left(1+\frac{2p_2}{p_2-p_1}\right)\la^2 s_1^{(0)}s_2^{(0)} \nonumber \\
&&+\frac{N-1}{6}\left(N+1-\frac{2p_2}{p_2-p_1}\right)\la^2 {s_2^{(0)}}^2\Biggr]
+\La \eeqq
we see that the ${\cal O}(s_i^{(0)})$-terms in eqn. (\ref{6.11}) agree with the 
logarithmic terms in the one-loop 
result (\ref{1.3}). The quadratic, $p_j$-dependent
terms in eqn. (\ref{6.11}) should be compared with the two-loop $\MS$ effective
potential \cite{cf1}
\beqq \label{6.12} V^{(\hbox{\tiny{$2$-loop}})}&=&
{\la {{\cal M}_1}^2\over8}\left(1-\log{{\cal M}_1\over{\mu^2}}\right)^2
+(N^2-1){\la {{\cal M}_2}^2\over{24}}
\left(1-\log{{\cal M}_2\over{\mu^2}}\right)^2 \nonumber \\
&+&(N-1){\la {\cal M}_1{\cal M}_2\over{12}}
\left(1-\log{{\cal M}_1\over{\m^2}}-\log{{\cal M}_2\over{\m^2}}+
\log{{\cal M}_1\over{\m^2}}\log{{\cal M}_2\over{\m^2}}\right) \nonumber \\
&-&{(\la\va)^2\over{12}}I({\cal M}_1,{\cal M}_1,{\cal M}_1)-
(N-1){(\la\va)^2\over{36}}I({\cal M}_2,{\cal M}_2,{\cal M}_1), \eeqq
where $I(x,y,z)$ 
is the general subtracted 
\lq\lq sunset'' vacuum integral discussed in appendix B.

Note that the sunset integrals do not contribute to the $m^4$-terms.
When comparing the $m^4$-terms in eqns. (\ref{6.11}) and (\ref{6.12})
it is easy to see
that they only agree for $p_1=1$ and $p_2=0$. Comparison of the $m^2\va^2$- and
$\va^4$-terms is more tricky due to the non-trivial sunset integrals.

We should decompose these integrals into logarithmic and non-logarithmic parts.
This is not too difficult for $I({\cal M}_1,{\cal M}_1,{\cal M}_1)$.
Unfortunately, the decomposition of $I({\cal M}_2,{\cal M}_2,{\cal M}_1)$ is
\sl not unique. \rm However, as discussed in appendix B it seems natural
to adopt the following one
\beqq \label{6.13}
I(x,y,z)&=&-{1\over2}\left[(y+z-x)\log{y\over{\mu^2}}\log{z\over{\m^2}}
+(z+x-y)\log{z\over{\mu^2}}\log{x\over{\m^2}}\right. \nonumber \\
&&\left.+(x+y-z)\log{x\over{\mu^2}}\log{y\over{\m^2}}\right]
+2x\log{x\over{\m^2}}+2y\log{y\over{\m^2}} \nonumber \\
&+&2z\log{z\over{\m^2}}+\hbox{\lq\lq non-logarithmic'' terms.}
\eeqq
Inserting eqn. (\ref{6.13}) into eqn. (\ref{6.12}) we see that the
$\va^4$-term 
agrees with the one in eqn. (\ref{6.11}). The $\va^2$-terms agree only if
$p_1=1$ and $p_2=0$.

An alternative check on eqn. (\ref{6.9}) is provided by the large $N$ 
limit. By construction our improved potential will agree with 
standard large $N$ results. Examining the next-to-large $N$ result \cite{root}
we have found that in the LL approximation the $m^4$-terms in eqn. (\ref{6.9})
and in the next-to-large $N$ limit expression only agree if $p_1=1$
and $p_2=0$.
To compare the $m^2\va^2$-terms we have again employed a \lq\lq natural''
decomposition of some integrals and once again 
agreement is achieved for $p_1=1$ and $p_2=0$. We remark that no other
choice of $p_j$ may be obtained by simply adopting a different
decomposition of the relevant integrals.
We have been unable to check the $\va^4$-terms since we do not know whether it
is possible to perform a \lq\lq natural'' decomposition of some of the 
contributing integrals. 

Thus, a comparison of our improved potential with the standard 
two-loop  and next-to-large $N$ potentials strongly indicates that
$p_1=1$ and $p_2=0$ is the appropriate choice.
This is gratifying, since for this choice one does not encounter the 
complex running parameters mentioned in the previous section.
Let us finally write down explicitly the two-scale improved potential in the
two-scale minimal subtraction scheme for this choice of $p_j$
\beqq \label{6.14}
V^{(0)}&=&{\la\va^4\over{24}}
\left(1-3\la s_1^{(0)}-\frac{N-1}{3}\la s_2^{(0)}\right)^{-1} \nonumber \\
&+&{m^2\va^2\over2}
\left(1-3\la s_1^{(0)}-\frac{N-1}{3}\la s_2^{(0)}\right)^{-\frac{1}{3}}
\left(1-{\frac{N+8}{3}\la s_2^{(0)}\over{1-3t}}\right)
^{-\frac{2}{3}\frac{N-1}{N+8}} \nonumber \\
&-&{m^4\over{2\la}}\left[
\left(1-3\la s_1^{(0)}-\frac{N-1}{3}\la s_2^{(0)}\right)^{\frac{1}{3}}
\left(1-{\frac{N+8}{3}\la s_2^{(0)}\over{1-3t}}\right)
^{-\frac{4}{3}\frac{N-1}{N+8}}-1\right] \nonumber \\
&+&2\,{N-1\over{N-4}}{m^4\over{\la}}(1-3t)^{\frac{1}{3}}
\left[\left(1-{\frac{N+8}{3}\la s_2^{(0)}\over{1-3t}}\right)
^{-\frac{N-4}{N+8}}-1\right]+\La. \eeqq  

For $t=0$ this result has already been obtained in a different way in
ref. \cite{cf2}. In the broken phase ($m^2<0$)
the tree-level minimum is at ${\cal M}_2=0$ or $s_2^{(0)}\ar-\infty$.
Hence, as we approach it $\log({\cal M}_2/{\cal M}_1)$ will become large.
If we are prepared to trust eqn. (\ref{6.14}) even in the \sl extreme \rm
case of the tree minimum itself an intriguing property emerges.

As long as $N>4$ the $\va^4$-and $m^2\va^2$-terms vanish and the
$m^4$-term converges to a finite value. As the slope
$\frac{dV^{(0)}}{ds_2^{(0)}}\tiny{(s_2^{(0)}\ar-\infty)}\searrow 0$
the EP takes its minimum in the broken phase at the tree-level value and
becomes complex for even smaller $\va^2$-values.
But for $1< N\leq4$ the 
$m^4$-term, and hence $V^{(0)}$, diverges to minus infinity.
This indicates that for these values of $N$ there is \sl no stable vacuum in
the broken phase. \rm Note especially that for $N=4$, ie. the SM scalar boson
content, the divergence is softer but still there, as the penultimate term in
eqn. (\ref{6.14})  becomes a logarithm
\beq V^{(0)}=.....-{m^4\over{2\la}}(1-3t)^{\frac{1}{3}}
\log\left(1-{4\la s_2^{(0)}\over{1-3t}}\right)+\La. \eeq

\section{NLO RG functions}

\paragraph{}
The LO results of the last two sections have already been obtained in a
less general form in ref. \cite{cf2} based on the use of the $\MS$ RG
(\ref{1.6}) and the conjecture that the correct boundary condition at
$\mu^2={\cal M}_2$
are given by the $N=1$ result (\ref{1.11}). But using those techniques
it appeared to be impossible to go beyond LO.
The finite renormalization (\ref{2.13}), introducing the appropriate number
of renormalization scales and the corresponding RG equations (\ref{2.15}),
allows us to overcome this problem in a systematic manner. To show the
strength of this technique 
we now determine the NLO RG functions and in the next section the
corresponding NLO running parameters.

To NLO eqn. (\ref{2.21}) yields
\beq [{\tilde\D}^{(1)},\D_i^{(0)}]+[{\tilde\D}^{(0)},\D_i^{(1)}]=0. \eeq
The corresponding equations for the various ${_i\be}^{(1)}_{.}$ are obtained
with the use of eqn. (\ref{2.25}). We now solve them in turn.

${_i\be}^{(1)}_\la$ is determined by
\beq D\,{_i\be}^{(1)}_\la
+{\tilde\be}^{(0)}_\la\frac{\pa}{\pa\la} {_i\be^{(1)}_\la}
-{_i\be}^{(1)}_\la\frac{\pa}{\pa\la}{\tilde\be}^{(0)}_\la 
+{\tilde\be}^{(1)}_\la\frac{\pa}{\pa\la} {_i\be^{(0)}_\la}
-{_i\be}^{(0)}_\la\frac{\pa}{\pa\la}{\tilde\be}^{(1)}_\la=0. \eeq
Proceeding in an analogous way as in obtaining the LO RG functions
in section 4 we easily obtain the solution
\beq \label{5.57} {_i\be^{(1)}_\la}=\la^3\al_i^{(1)}, \eeq
where
\beqq
& &\al_i^{(1)}(t)=a_{1i}^{(1)}+a_{2i}^{(1)} f^{-1}(t); \nonumber \\
& &\quad a_{1i}^{(1)}={\tilde A}^{(1)}a_{1i}^{(0)} \quad\mbox{and}\quad
{\tilde A}^{(1)}=\frac{{\tilde\al}^{(1)}}{{\tilde\al}^{(0)}}, \nonumber \\
& &\quad a_{2i}^{(1)}
=\frac{1}{{\tilde\al}^{(0)}}\left({\tilde\al}^{(0)}\al_i^{(1)}(0)-
\al_i^{(0)}(0){\tilde\al}^{(1)}\right). \eeqq

The equation for ${_i\be}^{(1)}_{m^2}$ is
\beq D\,{_i\be}^{(1)}_{m^2}
+{\tilde\be}^{(0)}_\la\frac{\pa}{\pa\la} {_i\be^{(1)}_{m^2}}
-{_i\be}^{(1)}_\la\frac{\pa}{\pa\la}{\tilde\be}^{(0)}_{m^2} 
+{\tilde\be}^{(1)}_\la\frac{\pa}{\pa\la} {_i\be^{(0)}_{m^2}}
-{_i\be}^{(0)}_\la\frac{\pa}{\pa\la}{\tilde\be}^{(1)}_{m^2}=0 \eeq
with the solution
\beq {_i\be}_{m^2}^{(1)}=m^2\la^2\be_i^{(1)}, \eeq
where
\beqq
& &\be_i^{(1)}(t)=b_{1i}^{(1)}+b_{2i}^{(1)}f^{-1}(t)+f^{-2}(t)
[b_{3i}^{(1)}+b_{4i}^{(1)} \log f(t)]; \nonumber \\
& &\quad b_{1i}^{(1)}={\tilde B}^{(1)}a_{1i}^{(0)} \quad\mbox{and}\quad
{\tilde B}^{(1)}=\frac{{\tilde\be}^{(1)}}{{\tilde\al}^{(0)}}, \quad
b_{2i}^{(1)}={\tilde B}^{(0)}a_{2i}^{(1)}, \nonumber \\
& &\quad b_{3i}^{(1)}=\frac{1}{{\tilde\al}^{(0)}}
\left({\tilde\al}^{(0)}\be_i^{(1)}(0)-
\al_i^{(0)}(0){\tilde\be}^{(1)}\right)-b_{2i}^{(1)}, 
\quad b_{4i}^{(1)}=-{\tilde A}^{(1)}b_{2i}^{(0)}. \eeqq

The equation for ${_i\be}^{(1)}_\La$ becomes quite involved
\beqq & &D\,{_i\be}^{(1)}_\La
+{\tilde\be}^{(0)}_\la\frac{\pa}{\pa\la} {_i\be^{(1)}_\La}
-{_i\be}^{(1)}_\la\frac{\pa}{\pa\la}{\tilde\be}^{(0)}_\La 
+{\tilde\be}^{(0)}_{m^2}\frac{\pa}{\pa m^2} {_i\be^{(1)}_\La}
-{_i\be}^{(1)}_{m^2}\frac{\pa}{\pa 
m^2}{\tilde\be}^{(0)}_\La \nonumber \\
& &+{\tilde\be}^{(1)}_\la\frac{\pa}{\pa\la} {_i\be^{(0)}_\La} 
-{_i\be}^{(0)}_\la\frac{\pa}{\pa\la}{\tilde\be}^{(1)}_\La
+{\tilde\be}^{(1)}_{m^2}\frac{\pa}{\pa m^2} {_i\be^{(0)}_\La}
-{_i\be}^{(0)}_{m^2}\frac{\pa}{\pa m^2}{\tilde\be}^{(1)}_\La=0. \eeqq
After some algebra we find the result
\beq {_i\be}_\La^{(1)}=m^4\la\ga_i^{(1)}, \eeq
where
\beqq
& &\ga_i^{(1)}(t)=c_{1i}^{(1)}+c_{2i}^{(1)}f^{-1}(t)
+f^{-2}(t)[c_{3i}^{(1)} +c_{4i}^{(1)}\log f(t)] \nonumber \\
& &\quad\quad\quad+f^{-2{\tilde B}^{(0)}}(t)\left[c_{5i}^{(1)}
+c_{6i}^{(1)}f^{-1}(t)+c_{7i}^{(1)}f^{-1}(t)\log f(t)\right]; \nonumber \\ 
& &\quad c_{1i}^{(1)}={\tilde C}^{(1)}a_{1i}^{(0)} \quad\mbox{and}\quad
{\tilde C}^{(1)}=\frac{{\tilde\ga}^{(1)}}{{\tilde\al}^{(0)}}, \nonumber \\
& &\quad c_{2i}^{(1)}={\tilde C}^{(0)}a_{2i}^{(1)}
+\left(\frac{{\tilde C}^{(1)}}{{\tilde B}^{(0)}}-
\frac{{\tilde C}^{(0)}}{{\tilde B}^{(0)}}
\frac{2{\tilde B}^{(1)}-{\tilde A}^{(1)}}
{2{\tilde B}^{(0)}-1}\right)b_{2i}^{(0)}, \nonumber \\
& &\quad c_{3i}^{(1)}=\frac{2{\tilde C}^{(0)}}
{2{\tilde B}^{(0)}-1}b_{3i}^{(1)}, \quad
c_{4i}^{(1)}=-{\tilde A}^{(1)}c_{2i}^{(0)}, \nonumber \\
& &\quad c_{5i}^{(1)}=2\left({\tilde A}^{(1)}{\tilde B}^{(0)}-
{\tilde B}^{(1)}\right)c_{3i}^{(0)}, \nonumber \\
& &\quad c_{6i}^{(1)}=\ga_i^{(1)}(0)-c_{1i}^{(1)}-c_{2i}^{(1)}
-c_{3i}^{(1)}-c_{5i}^{(1)}, 
\quad c_{7i}^{(1)}=-2{\tilde A}^{(1)}{\tilde B}^{(0)}c_{3i}^{(0)}. \eeqq

To NLO the anomalous dimension is non-trivial and we have to determine
${_i\be}^{(1)}_\va$ from
\beq D\,{_i\be}^{(1)}_\va
+{\tilde\be}^{(0)}_\la\frac{\pa}{\pa\la} {_i\be^{(1)}_\va}
-{_i\be}^{(1)}_\la\frac{\pa}{\pa\la}{\tilde\be}^{(0)}_\va 
+{\tilde\be}^{(1)}_\la\frac{\pa}{\pa\la} {_i\be^{(0)}_\va}
-{_i\be}^{(0)}_\la\frac{\pa}{\pa\la}{\tilde\be}^{(1)}_\va=0. \eeq
The solution is easily obtained
\beq {_i\be^{(1)}_\va}=\la^2\de_i^{(1)}, \eeq
where
\beqq 
& &\de_i^{(1)}(t)=d_{1i}^{(1)}+d_{2i}^{(1)}f^{-2}(t); \nonumber \\
& &\quad d_{1i}^{(1)}={\tilde D}^{(1)}a_{1i}^{(0)} \quad\mbox{and}\quad
{\tilde D}^{(1)}=\frac{{\tilde\de}^{(1)}}{{\tilde\al}^{(0)}}, \nonumber \\
& &\quad d_{2i}^{(1)}=\frac{1}{{\tilde\al}^{(0)}}\left({\tilde\al}^{(0)}
\de_i^{(1)}(0)-\al_i^{(0)}(0){\tilde\de}^{(1)}\right). \eeqq

So far we have not specified the values of the NLO boundary 
constants $\al_i^{(1)}(0)$, $\be_i^{(1)}(0)$, $\ga_i^{(1)}(0)$ and
$\de_i^{(1)}(0)$. At LO the relevant constants were completely determined by the
single-scale limit conditions following from requirements ii) and iii).
Unfortunately they do not anymore \sl uniquely 
\rm fix the NLO constants. For suppose we expand
$\al_i^{(1)}(0)$, $\be_i^{(1)}(0)$, $\ga_i^{(1)}(0)$ and
$\de_i^{(1)}(0)$ in powers of $(N-1)$. Then 
the large $N$ limit condition forbids any terms proportional to
$(N-1)^2$ and higher powers of $(N-1)$ \cite{largen}, and the $N=1$ limit 
condition fixes the contributions proportional to $(N-1)^0$.
However, these limits tell us nothing about NLO terms proportional 
to $(N-1)$. Of course, we still have the condition that the sums of
the two sets of RG functions at $t=0$ are just the usual $\MS$ RG
functions, ie. ${}_1\be^{(1)}_.(t=0)+{}_2\be^{(1)}_.(t=0)=
\be^{(\hbox{\tiny{$2$-loop}})}_{.\, ,\MSf}$. In $\MS$
$\be^{(\hbox{\tiny{$2$-loop}})}_{\La\, ,\MSf}=0$
and the other two-loop beta functions can be found eg. in ref. \cite{cf1}.
Putting all this together we have
\beqq \label{7.14} \al_1^{(1)}(0)&=&-\hbox{${17\over3}$}-[1+q_1](N-1),\quad
\al_2^{(1)}(0)=q_1(N-1), \nonumber \\
\be_1^{(1)}(0)&=&-\hbox{${5\over6}$}-[\hbox{${5\over{18}}$}+q_2](N-1),\quad
\be_2^{(1)}(0)=q_2(N-1), \nonumber \\
\ga_1^{(1)}(0)&=&q_3(N-1),\quad \ga_2^{(1)}(0)=-q_3(N-1), \nonumber \\
\de_1^{(1)}(0)&=&\hbox{${1\over{12}}$}+[\hbox{${1\over{36}}$}+q_4](N-1),\quad
\de_2^{(1)}(0)=-q_4(N-1),\eeqq
where $q_j$ are real numbers which are independent of $N$. We shall comment
further on sensible choices for $q_j$ in the discussion of the NLO effective
potential in section 9.

\section{NLO running two-scale parameters}

\paragraph{}
Using the LO results and the set of RG functions obtained in the last section
we now calculate the NLO running two-scale parameters, which will be 
used to construct the NLO effective potential.

The equation for the next-to-leading order running two-scale coupling is
\beq \frac{d\la^{(1)}}{d s_i}=2\la^{(0)}\al_i^{(0)}\la^{(1)}
+{\la^{(0)}}^3\al_i^{(1)}. \eeq
With the use of the results (\ref{4.42}) for $\la^{(0)}$ and (\ref{5.57})
for $\al_i^{(1)}$ we may integrate this equation and find
\beq \la^{(1)}(s_i)=\la^{(0)}(s_i)^2
\log\left(\left(\frac{\la^{(0)}(s_i)}{\la}\right)^{A^{(1)}}
\left(\frac{f^{(0)}(s_i)}{f}\right)
^{{\tilde A}^{(1)}-A^{(1)}}\right).\eeq
Above $A^{(1)}=\frac{\al_1^{(1)}+\al_2^{(1)}}{\al_1^{(0)}+\al_2^{(0)}}$.

Turning to the NLO running mass we have to solve
\beq \frac{d {m^2}^{(1)}}{d s_i}
=\la^{(0)}\be_i^{(0)}{m^2}^{(1)}+{m^2}^{(0)}\left(\be_i^{(0)}\la^{(1)}
+\la^{(0)}\frac{\pa\be_i^{(0)}}{\pa\la}\la^{(1)}
+{\la^{(0)}}^2 \be_i^{(1)}\right). \eeq
The integration of this equation is quite involved and yields
\beqq {m^2}^{(1)}(s_i)&=&{m^2}^{(0)}(s_i)\left[
M_1^{(1)}\left[\la^{(0)}(s_i)-\la\right]
+M_2^{(1)}\left[\frac{\la^{(0)}(s_i)}{f^{(0)}(s_i)}-
\frac{\la}{f}\right]\right. \nonumber \\
&&+\la^{(0)}(s_i)\left[M_3^{(1)}\log\left(\frac{f^{(0)}(s_i)}{f}\right)
+M_4^{(1)}\log\left(\frac{\la^{(0)}(s_i)}{\la}\right)\right] \nonumber \\
&&\left.+M_5^{(1)}\frac{\la^{(0)}(s_i)}{f^{(0)}(s_i)}
\log\left(\frac{f^{(0)}(s_i)}{f}\right)
\left(\frac{\la^{(0)}(s_i)}{\la}\right)^{-1}\right], \eeqq
where
\beqq M_1^{(1)}
&=&{\tilde B}^{(1)}-{\tilde B}^{(0)}{\tilde A}^{(1)}, \nonumber \\
M_2^{(1)}&=&B^{(1)}-B^{(0)}A^{(1)}-M_1^{(1)}
+{\tilde A}^{(1)}({\tilde B}^{(0)}-B^{(0)})\log f, \nonumber \\
M_3^{(1)}&=&{\tilde B}^{(0)}({\tilde A}^{(1)}-A^{(1)}), \quad
M_4^{(1)}={\tilde B}^{(0)} A^{(1)}, \nonumber \\
M_5^{(1)}&=&M_4^{(1)}-B^{(0)}A^{(1)}. \eeqq
Above $B^{(1)}=\frac{\be_1^{(1)}+\be_2^{(1)}}{\al_1^{(0)}+\al_2^{(0)}}$.

The NLO running cosmological constant is determined by
\beq \frac{d \La^{(1)}}{d s_i}
=2{m^2}^{(0)}\ga_i^{(0)}{m^2}^{(1)}+({m^2}^{(0)})^2
\Big(\frac{\pa\ga_i^{(0)}}{\pa\la}\la^{(1)}
+\la^{(0)}\ga_i^{(1)}\Big). \eeq
With the use of the various results above we obtain after a tedious computation
\beqq \label{8.7}
\La^{(1)}(s_i)&=&\la\;L_1^{(1)}\left[\frac{({m^2}^{(0)}(s_i))^2}{\la^{(0)}(s_i)}
-\frac{m^4}{\la}\right] \nonumber \\
&+&\la\;L_2^{(1)}\left[\frac{({m^2}^{(0)}(s_i))^2}{\la^{(0)}(s_i)}
f^{(0)}(s_i)^{1-2{\tilde B}^{(0)}}-
\frac{m^4}{\la}f^{1-2{\tilde B}^{(0)}}\right] \nonumber \\
&+&L_3^{(1)}\left[({m^2}^{(0)}(s_i))^2-m^4\right]
+L_4^{(1)}\left[\frac{({m^2}^{(0)}(s_i))^2}{f^{(0)}(s_i)}
-\frac{m^4}{f}\right] \nonumber \\ 
&+&L_5^{(1)}\left[\frac{({m^2}^{(0)}(s_i))^2}{f^{(0)}(s_i)^{2{\tilde B}^{(0)}}}
-\frac{m^4}{f^{2{\tilde B}^{(0)}}}\right] \nonumber \\ 
&+&({m^2}^{(0)}(s_i))^2\left[L_6^{(1)}\log\left(\frac{f^{(0)}(s_i)}{f}\right)
+L_7^{(1)}\log\left(\frac{\la^{(0)}(s_i)}{\la}\right)\right] \nonumber \\
&+&L_8^{(1)}\frac{({m^2}^{(0)}(s_i))^2}{f^{(0)}(s_i)}
\log\left(\frac{f^{(0)}(s_i)}{f}\right)
\left(\frac{\la^{(0)}(s_i)}{\la}\right)^{-1} \nonumber \\
&+&L_9^{(1)}\frac{({m^2}^{(0)}(s_i))^2}{f^{(0)}(s_i)^{2{\tilde B}^{(0)}}}
\log\left(\frac{f^{(0)}(s_i)}{f}\right)
\left(\frac{\la^{(0)}(s_i)}{\la}\right)^{-1}, \eeqq
where
\beqq L_1^{(1)}&=&-2\left(M_1^{(1)}+\frac{1}{f} M_2^{(1)}\right)
L_1^{(0)}, \quad 
L_2^{(1)}=-2\left(M_1^{(1)}+\frac{1}{f} M_2^{(1)}\right)
L_2^{(0)}, \nonumber \\
L_3^{(1)}&=&\frac{{\tilde C}^{(0)}}{2{\tilde B}^{(0)}}
\left(2M_1^{(1)}+\frac{{\tilde C}^{(1)}}{{\tilde C}^{(0)}}-{\tilde A}^{(1)}
\right), \quad
L_4^{(1)}=\frac{2{\tilde C}^{(0)}}{2{\tilde B}^{(0)}-1}M_2^{(1)} \nonumber \\
L_5^{(1)}&=&\frac{1}{2B^{(0)}}\left(C^{(1)}-C^{(0)}A^{(1)}\right)
+\frac{C^{(0)}}{B^{(0)}}\left(B^{(1)}-B^{(0)}A^{(1)}\right)
-\frac{{\tilde C}^{(1)}}{2{\tilde B}^{(0)}} \nonumber \\
&-&\frac{{\tilde C}^{(0)}}{2{\tilde B}^{(0)}-1}\left(2B^{(1)}-2B^{(0)}A^{(1)}
+\frac{{\tilde A}^{(1)}}{2{\tilde B}^{(0)}}
-\frac{{\tilde B}^{(1)}}{{\tilde B}^{(0)}}\right) \nonumber \\
&+&{\tilde A}^{(1)}\left({\tilde C}^{(0)}-C^{(0)}
-\frac{2{\tilde C}^{(0)}}{2{\tilde B}^{(0)}-1}({\tilde B}^{(0)}-B^{(0)})
\right) \log f, \nonumber \\
L_6^{(1)}&=&{\tilde C}^{(0)}{\tilde A}^{(1)}-L_7^{(1)}, \quad
L_7^{(1)}={\tilde C}^{(0)}A^{(1)}, \nonumber \\
L_8^{(1)}&=&\frac{2{\tilde C}^{(0)}}{2{\tilde B}^{(0)}-1}M_5^{(1)}, \quad
L_9^{(1)}=-C^{(0)}A^{(1)}+L_7^{(1)}-L_8^{(1)}. \eeqq
Above $C^{(1)}=\frac{\ga_1^{(1)}+\ga_2^{(1)}}{\al_1^{(0)}+\al_2^{(0)}}$.
We remark that most 
individual integrals occurring in the computation of not only
$\La^{(1)}$ but also $\La^{(0)}$ and ${m^2}^{(1)}$ yield hypergeometric
functions and that only the respective sums of those are again expressible in
terms of elementary functions as given above.

Finally we determine the non-trivial NLO running of $\va(s_i)$
\beq \frac{d\va^{(1)}}{d s_i}=-\va^{(0)}{\la^{(0)}}^2\de_i^{(1)},
\quad \va(s_i=0)=\va. \eeq
The integration of this equation is straightforward and yields
\beqq \va^{(1)}(s_i)&=&-\va{\tilde D}^{(1)}
\left[\la^{(0)}(s_i)-\la\right] \nonumber \\
&+&\va({\tilde D}^{(1)}-D^{(1)})\left[\frac{\la^{(0)}(s_i)}{f^{(0)}(s_i)}
-\frac{\la}{f}\right], \eeqq
where $D^{(1)}=\frac{\de_1^{(0)}+\de_2^{(0)}}{\al_1^{(0)}+\al_2^{(0)}}$.

It is easy to see that $\la^{(1)}$, ${m^2}^{(1)}$ and $\va^{(1)}$ vanish
for $N>1$ in the limit of one $s_i\ar -\infty$ while holding the other fixed.
$\La^{(1)}$ will tend to a finite value in this limit
only for $N>4$. However, it will diverge for $1<N\leq 4$
if $p_1=1$ and $s_2\ar -\infty$ with the same rate as $\La^{(0)}$ due
to the first two terms in (\ref{8.7}).

\section{NLO RG improved potential}

\paragraph{}
It is straightforward to extract the two-scale NLO potential from the standard
perturbative boundary condition eqn. (\ref{6.8})
\beqq \label{9.1}
V^{(1)}(\la,...;\ka_i)&=&{\la^{(1)}(s_i^{(0)})\over{24}}\va^4
+{\la^{(0)}(s_i^{(0)})\over{6}}\,\va^3\,\va^{(1)}(s_i^{(0)}) \nonumber \\
&+&{1\over2}{m^2}^{(1)}(s_i^{(0)})\va^2
+{m^2}^{(0)}(s_i^{(0)})\,
\va\,\va^{(1)}(s_i^{(0)})+\La^{(1)}(s_i^{(0)}) \nonumber \\
&+&\sum_{i=1}^2\biggr[{_i\be}^{(0)}_{\la}(s_j^{(0)}){\va^4\over{24}}
+{_i\be}^{(0)}_{m^2}
(s_j^{(0)}){\va^2\over2}+{_i\be}^{(0)}_{\La}(s_j^{(0)})\biggl]
s_i^{(1)}(\la,...;s_i^{(0)}) \nonumber \\
&+&{3\over 2}\left(\frac{{{\cal M}^{(0)}_1(s_i^{(0)})}^2}{4}
+(N-1)\frac{{{\cal M}^{(0)}_2(s_i^{(0)})}^2}{4}\right). \eeqq
The different contributions come from the expansion of the running two-scale
parameters, from the expansion of their $s_i$-dependence and from
the explicit one-loop term in (\ref{6.8}). In practice, we immediately set
$p_1=1$ and $p_2=0$ as has been done in the LO result.

Next, we fix the values of $q_j$ used to parameterize the NLO boundary
functions in eqn. (\ref{7.14}) by comparing the $q_j$-dependent NLO potential
\sl and the NLO 
$Z(\va)^{(1)}$-function \rm with the corresponding standard two-loop
results. This immediately fixes $q_3=0$ and hence $\ga_i^{(1)}(0)=0$.
The value of $q_4$ depends on how we decompose the two-loop integral $J$
for $Z(\va)^{(\hbox{\tiny{$2$-loop}})}$
into its logarithmic and non-logarithmic pieces.

In order to determine the \lq\lq natural'' decomposition of this
integral it is helpful to consider the \sl general \rm integral
$J(x,y,z)$ as given in appendix B. It is symmetric in
$x,y,z$. Accordingly, a natural decomposition should respect this
property. In fact, there is only one decomposition which does this
\beq
J(x,y,z)\propto\log{x\over{\mu^2}}+\log{y\over{\mu^2}}
+\log{z\over{\mu^2}}+\hbox{\lq\lq non-logarithmic'' terms.} \eeq
We are interested in the case $J({\cal M}_2,{\cal M}_2,{\cal M}_1)$ and
so we choose the coefficient of the $\log({\cal M}_2/\mu^2)$-term in
$J({\cal M}_2,{\cal M}_2,{\cal M}_1)$ to be twice that of the
$\log({\cal M}_1/\mu^2)$-term.
This implies that the coefficient of $(N-1)$ in $\delta_2^{(1)}(0)$ must
be twice the coefficient of $(N-1)$ in $\delta_1^{(1)}(0)$ or
$q_4=-\hbox{${1\over{54}}$}$.

To determine $q_1$ and $q_2$ we need the subleading logarithms 
in $I({\cal M}_2,{\cal M}_2,{\cal M}_1)$. Using the decomposition
(\ref{6.12}) yields 
$q_1=-\hbox{${10\over{27}}$}$ and $q_2=-\hbox{${5\over{27}}$}$.
Putting this all together, the complete set of boundary functions are
\beqq \al_1^{(1)}(0)&=&-\hbox{${17\over3}$}-\hbox{${17\over{27}}$}(N-1),\quad
\al_2^{(1)}(0)=-\hbox{${10\over{27}}$}(N-1), \nonumber\\
\be_1^{(1)}(0)&=&-\hbox{${5\over6}$}-\hbox{${5\over{54}}$}(N-1),\quad
\be_2^{(1)}(0)=-\hbox{${5\over{27}}$}(N-1), \nonumber \\
\ga_1^{(1)}(0)&=&0,\quad \ga_2^{(1)}(0)=0, \nonumber \\
\de_1^{(1)}(0)&=&\hbox{${1\over{12}}$}+\hbox{${1\over{108}}$}(N-1),\quad
\de_2^{(1)}(0)=\hbox{${1\over{54}}$}(N-1).\eeqq

The behaviour of the NLO contribution is of most interest
around the broken phase tree-level minimum, where ${\cal M}_2=0$
or $s_2^{(0)}\ar-\infty$.
As in the LO case all the terms in eqn. (\ref{9.1}) will vanish or converge
to a finite limit if $N>4$. But for $1<N\leq 4$ $\La^{(1)}$ and
${}_2\be_\La^{(0)}\cdot s_2^{(1)}$ will diverge.
It is easy to check that they diverge at the same rate as $\La^{(0)}$ in the
LO analysis. However, as the NLO divergence is suppressed by a factor
$\frac{\hbar\la}{(4\pi)^2} \ll 1$ qualitatively nothing will change.

\section{The relevance of $N=2$}

\paragraph{}
From diagrammatic considerations we would expect the $m^4$-term in the
RG improved potential to have a certain exchange symmetry in the $N=2$ case.
For the case $N=2$ the $m^4$-contributions are invariant
under the exchange of Higgs- and Goldstone-lines. We would therefore
expect that for $\ka_1=\ka_2=\mu$ the $m^4$-terms in eqn. (\ref{6.14}) should
be symmetric in $s_1^{(0)}$ and $s_2^{(0)}$. A glance at eqn. (\ref{6.14})
in this case,
\beqq \label{10.1}
V^{(0)}&=&-{m^4\over {2\la}}\left[
(1-3\la s_1^{(0)}-\th\la s_2^{(0)})^{{1\over 3}}
(1-\hbox{${10\over3}$}\la s_2^{(0)})^{-{2\over 15}}
+2(1-\hbox{${10\over3}$}\la s_2^{(0)})^{{1\over 5}}-3\right] \nonumber \\
&+&\hbox{other terms}, \eeqq
clearly shows that the $m^4$-term is \sl not \rm symmetric in $s_1^{(0)}$
and $s_2^{(0)}$. We find it somewhat disturbing that our approximation 
scheme does not respect this symmetry.

We know from section 6 that
eqn. (\ref{10.1}) matches standard perturbation theory through to two 
loops. Therefore, 
this $s_1^{(0)}\leftrightarrow s_2^{(0)}$ symmetry must go down
\sl beyond \rm the two-loop level. Expanding eqn. (\ref{10.1}) in powers of
$s_1^{(0)}$ and $s_2^{(0)}$ up to ${\cal O}(\la^5)$
\beqq
V^{(0)}&=&-{m^4\over{2}}\Biggl[
s_1^{(0)}+s_2^{(0)}+\la\left({s_1^{(0)}}^2
+\hbox{${2\over3}$}s_1^{(0)}s_2^{(0)}+{s_2^{(0)}}^2\right) \nonumber \\
&&+\la^2\left(\hbox{${5\over3}$}{s_1^{(0)}}^3
+{s_1^{(0)}}^2s_2^{(0)}+s_1^{(0)}{s_2^{(0)}}^2
+\hbox{${5\over3}$}{s_2^{(0)}}^3\right) \nonumber \\
&&+\la^3\left(\hbox{${10\over3}$}{s_1^{(0)}}^4
+\hbox{${20\over9}$}{s_1^{(0)}}^3s_2^{(0)}
+\hbox{${4\over3}$}{s_1^{(0)}}^2{s_2^{(0)}}^2
+\hbox{${20\over9}$}s_1^{(0)}{s_2^{(0)}}^3
+\hbox{${10\over3}$}{s_2^{(0)}}^4\right) \nonumber \\
&&+\la^4\biggl(\hbox{${22\over3}$}{s_1^{(0)}}^5
+\hbox{${50\over9}$}{s_1^{(0)}}^4s_2^{(0)}
+\hbox{${80\over{27}}$}{s_1^{(0)}}^3{s_2^{(0)}}^2 \nonumber \\
&&\quad+\hbox{${20\over9}$}{s_1^{(0)}}^2{s_2^{(0)}}^3
+\hbox{${178\over{27}}$}s_1^{(0)}{s_2^{(0)}}^4
+\hbox{${986\over{135}}$}{s_2^{(0)}}^5\biggr)\Biggr] \eeqq
we see that 
the $s_1^{(0)}\leftrightarrow s_2^{(0)}$ symmetry \sl survives \rm at
three and four loops, but breaks  down at \sl five \rm loops.
So we see that the failure of our approximation to observe
it only appears at quite a high order in perturbation theory. We
are unable to explain this phenomenon further.

\section{Conclusions}

\paragraph{}
In order to deal systematically with the two-scale problem arising in
the analysis of the effective potential in the $O(N)$-symmetric
$\phi^4$-theory we have introduced a generalization of $\MS$. At
each order in a $\MS$ loop-expansion we have performed a finite
renormalization to switch over to a new \lq\lq minimal two-scale
subtraction scheme'' allowing for two renormalization scales
$\ka_i$ corresponding to the two generic scales in the problem.
The $\MS$ RG functions and $\MS$ RGE then split into two minimal
two-scale subtraction scheme \lq\lq partial'' RG functions and
two \lq\lq partial'' RGE's. The respective integrability condition
inevitably imposes a dependence of the partial RG functions on the
renormalization scale ratio $\ka_2/\ka_1$. Supplementing the integrability
with an appropriate subsidiary condition we have been able to determine
this dependence to all orders in the scale ratio and have obtained a
trustworthy set of LO and NLO two-scale subtraction scheme RG functions.
With the use of the two \lq\lq partial'' RGE's we have then turned those
into LO and NLO running two-scale parameters exhibiting features similar
to the $\MS$ couplings such as a Landau pole now in both scaling channels.
Using standard perturbative boundary conditions, which become applicable
in the minimal two-scale subtraction scheme, we have calculated the
effective potential in this scheme to LO and NLO. To fix  the
remaining renormalization freedom we have compared our results with two-loop
and next-to-large $N$ limit $\MS$ calculations. As a main result we have
found in both LO and NLO that for $1<N\le 4$ there is no stable
vacuum in the broken phase.

The vacuum instability in the broken phase of the $O(N)$-model
raises immediately the  possibility of a similar
outcome in a multi-scale analysis of the SM effective potential.
As the method outlined generalizes naturally to problems with more than two
scales we are in a position to investigate systematically the different
possible scenarios. Before turning to the  SM itself it proves useful
thereby to study the effects of adding either fermions as in a
Yukawa-type model or gauging the simplest case of $N=2$ as in the Abelian-Higgs
model. The Yukawa 
case will either be a two- or three-scale problem, depending on
whether one includes Goldstone bosons or not. The Abelian-Higgs model in
the Landau gauge will be a three-scale problem to which
the methods in this paper are easily extended. Now one has \sl three \rm
integrability conditions $[{\cal D}_i,{\cal D}_j]=0$ and one must impose
three independent subsidiary conditions analogous to
$[\ka_1\pa/\pa \ka_1,{\cal D}_1]=0$ which we used in our $O(N)$-model
analysis. Note that for the general $n$-scale problem one would have
$\ha n(n-1)$ integrability conditions which should be supplemented by
$\ha n(n-1)$ subsidiary conditions. The question of whether fermions
or gauge fields may stabilize the effective potential for small $N$
in a full multi-scale analysis is under investigation.

We do not see any fundamental problem in applying the framework presented
here to multi-scale computations at finite temperature, to the analysis of the
multi-scale EP in supersymmetric extensions of the SM or to a full
multi-scale treatment of DIS problems in the regions of very large
or small $x_B$. The necessary adaptions are under investigation.

\section*{Acknowledgments}

\paragraph{}
C. W. has been partially supported by Schweizerischer Nationalfonds. We thank
D. O'Connor, L. O'Raifeartaigh, I. Sachs and C. R. Stephens for helpful 
comments.

\appendix

\section{Values of various constants}

\paragraph{}
Here, we give the values of various constants appearing in the paper. We quote
them for the choice $p_1=1$ and $p_2=0$.

\beqq
B^{(0)}&=&\frac{N+2}{N+8}, \quad
C^{(0)}=\frac{3N}{2(N+8)} \\ \nonumber \\
{\tilde B}^{(0)}&=&\frac{1}{3}, \quad
{\tilde C}^{(0)}=\frac{1}{6} \\ \nonumber \\
A^{(1)}&=&-\frac{3N+14}{N+8}, \quad
B^{(1)}=-\frac{5(N+2)}{6(N+8)}, \nonumber \\
C^{(1)}&=&0, \quad
D^{(1)}=\frac{N+2}{12(N+8)} \\ \nonumber \\
{\tilde A}^{(1)}&=&-\frac{17(N+8)}{81}, \quad
{\tilde B}^{(1)}=-\frac{5(N+8)}{162}, \nonumber \\
{\tilde C}^{(1)}&=&0, \quad
{\tilde D}^{(1)}=\frac{N+8}{324} \\ \nonumber \\
L_1^{(0)}&=&-\frac{1}{2}, \quad
L_2^{(0)}=\frac{2(N-1)}{N-4} \\ \nonumber \\
M_1^{(1)}&=&\frac{19(N+8)}{486}, \nonumber \\
M_2^{(1)}&=&-\frac{(N-1)(19N^2-578N-2600)}{486(N+8)^2}
+\frac{(N-1)(34N^2+544N+2178)}{243(N+8)^2}\log f, \nonumber \\
M_3^{(1)}&=&-\frac{(N-1)(17N+46)}{243(N+8)}, \quad
M_4^{(1)}=-\frac{3N+14}{3(N+8)}, \nonumber \\
M_5^{(1)}&=&\frac{2(N-1)(3N+14)}{3(N+8)^2} \\ \nonumber \\
L_1^{(1)}&=&\frac{19(N+8)}{486}
-\frac{(N-1)(19N^2-578N-2600)}{486(N+8)^2 f} \nonumber \\
&+&\frac{(N-1)(34N^2+544N+2178)}{243(N+8)^2 f}\log f, \nonumber \\
L_2^{(1)}&=&-\frac{38(N-1)(N+8)}{243(N-4)}
+\frac{2(N-1)^2(19N^2-578N-2600)}{243(N-4)(N+8)^2 f} \nonumber \\
&-&\frac{4(N-1)^2(34N^2+544N+2178)}{243(N-4)(N+8)^2 f}\log f, \quad
L_3^{(1)}=\frac{35(N+8)}{486}, \nonumber \\
L_4^{(1)}&=&\frac{(N-1)(19N^2-578N-2600)}{486(N+8)^2}
-\frac{(N-1)(34N^2+544N+2178)}{243(N+8)^2}\log f, \nonumber \\
L_5^{(1)}&=&-\frac{(N-1)(N^3-42N^2-360N-760)}{9(N+2)(N+8)^2}
+\frac{34(N-1)}{81}\log f, \nonumber \\
L_6^{(1)}&=&-\frac{(N-1)(17N+46)}{486(N+8)}, \quad
L_7^{(1)}=-\frac{3N+14}{6(N+8)}, \nonumber \\
L_8^{(1)}&=&-\frac{2(N-1)(3N+14)}{3(N+8)^2}, \quad
L_9^{(1)}=\frac{2(N-1)(3N+14)}{(N+8)^2} \eeqq

\section{The integrals $I$ and $J$}

\paragraph{}
Here, we list some useful formulae regarding the two-loop integrals $I$
and $J$. The general unsubtracted scalar sunset integral in $D$ dimensions
is defined as
\beq
I_D(x,y,z)=\int{d^Dk\over{(2\pi)^D}} {d^Dl\over{(2\pi)^D}}
{1\over{(k^2+x)(l^2+z)((k+l)^2+z)}}. \eeq
A full calculation of this integral is rather involved \cite{int}.
However, there is a formula in ref. \cite{dt} which
nicely splits the integral into a very simple, for $D=4$ divergent 
expression plus a finite term which is proportional to $I_{D-2}(x,y,z)$,
ie. the same integral in two lower dimensions.
\beqq
I_D(x,y,z)&=&(4\pi)^{-D}{\Ga^2(2-\ha D)\over{(D-2)(D-3)}}\Biggl[
(x-y-z)(yz)^{\ha D-2} \nonumber \\
&&+(y-z-x)(zx)^{\ha D-2}+(z-x-y)(xy)^{\ha D-2}\Biggr] \nonumber \\
&-&(4\pi)^{-2}(x^2+y^2+z^2-2xy-2yz-2zx)\;I_{D-2}(x,y,z). \eeqq
Since the last term is finite we regard it as a \lq\lq non-logarithmic''
term and ascribe the logarithmic terms purely to the simple, divergent
piece. The renormalized $I(x,y,z)$ referred to in the text is then
given as
\beqq
I(x,y,z)&=&\hbox{FP}\biggl[
(4\pi e^{-\gamma}\mu^2)^{2\epsilon}\Bigl(I_{4-2\epsilon}(x,y,z)
\nonumber \\
& &-\frac{1}{(4\pi)^2\epsilon}
\bigl(K_{4-2\epsilon}(x)+K_{4-2\epsilon}(y)+K_{4-2\epsilon}(z)\bigr)
\Bigr)\biggr]
\eeqq
where FP denotes the finite part, $\gamma$ is Euler's constant and
\beq
K_D(x)=\int{d^Dk\over{(2\pi)^D}}{1\over{k^2+x}}.
\eeq
The $K_D$-terms in eqn. (B.3) are due to the subtraction of
one-loop sub-divergences.

The unsubtracted $J_D(x,y,z)$ is defined as
\beq \label{JD}
J_D(x,y,z)=\left.{\partial\over{\partial p^2}}
\int{d^Dk\over{(2\pi)^D}}{d^Dl\over{(2\pi)^D}}
{1\over{(k^2+x)(l^2+y)((k+l+p)^2+z)}}\right|_{p^2=0}.
\eeq
The renormalized 
$J(x,y,z)$ which enters into $Z(\va)^{(\hbox{\tiny{$2$-loop}})}$
is simply
\beq
J(x,y,z)=\hbox{FP}\left[(4\pi e^{-\gamma}\mu^2)^{2\epsilon}
J_{4-2\epsilon}(x,y,z)\right].
\eeq
Above, the $x$, 
$y$, $z$ are the $(\hbox{masses})^2$ on the three internal lines.

\end{document}